\documentstyle[pre,preprint,aps]{revtex}
\tightenlines
\newcommand{\siml}{\stackrel{<}{\sim}}
\newcommand{\simg}{\stackrel{>}{\sim}}

\begin{document}
\draft

\title{
An augmented moment method for 
stochastic ensembles \\
with delayed couplings: 
II. FitzHugh-Nagumo model
}
\author{
Hideo Hasegawa
\footnote{e-mail:  hasegawa@u-gakugei.ac.jp}
}
\address{
Department of Physics, Tokyo Gakugei University,
Koganei, Tokyo 184-8501, Japan
}
\date{\today}
\maketitle
\begin{abstract}
Dynamics of FitzHugh-Nagumo (FN) 
neuron ensembles with time-delayed couplings
subject to white noises,
has been studied by using both
direct simulations and 
a semi-analytical augmented moment method (AMM)
which has been proposed 
in a preceding paper
[H. Hasegawa, Phys. Rev E {\bf xx}, yyyy (2004)].
For $N$-unit FN neuron ensembles, AMM transforms original
$2N$-dimensional {\it stochastic} delay differential equations 
(SDDEs) to infinite-dimensional {\it deterministic} DEs
for means and correlation functions 
of local and global variables.
Infinite-order recursive DEs are terminated 
at the finite level $m$
in the level-$m$ AMM (AMM$m$), yielding
$8(m+1)$-dimensional deterministic DEs.
When a single spike is applied, the oscillation may be induced
when parameters of coupling strength, delay, noise intensity 
and/or ensemble size are appropriate. 
Effects of these parameters 
on the emergence of the oscillation
and on the synchronization in FN neuron ensembles have been studied.
The synchronization shows the {\it fluctuation-induced} enhancement
at the transition between non-oscillating and oscillating states.
Results calculated by AMM5 
are in fairly good agreement with those obtained 
by direct simulations.

\end{abstract}

\noindent
\vspace{0.5cm}
%
\section{INTRODUCTION}


There have been many studies on effects
of noises in dynamical systems with delays.
Complex behavior due to noise and delay is found in many
systems such as biological systems, signal transmissions,
electrical circuits and lasers.
Systems with both noises and delay are commonly described by
stochastic delay differential equations (SDDEs).
In recent years, linear SDDEs of Langevin equation
are beginning to gain much attention
\cite{Kuchler92}-\cite{Frank03}.
The parameter range for the stationary solutions of
the Langevin equation has been examined
with the use of the step by step method \cite{Kuchler92},
the moment mothod \cite{Mackey95}
and the Fokker-Planck equation (FPE) method \cite{Duil99}\cite{Ohira00}.

When we pay our attention to living brains, various kinds of
noises are reported to be ubiquitous.
A study on noise effects has been one of major recent
topics in neuronal systems.
It has been shown that the response of neurons
may be improved by background noises.
The typical example is the stochastic resonance in which
weak noises enhance the transmission of signals
with the subthreshold level.
The transmission delay is inherent because the speed of 
spikes propagating through axons is finite. 
Conduction velocity ranges from 20 to 60 m/s,
leading to non-negligible transmission times from
milliseconds to hundreds milliseconds.   
Although an importance of effects of delay 
has been not so recognized as that of noises, 
there is an increasing interest in 
the complex behavior of time delays,
whose effects have been investigated by using 
integrate-and-fire (IF) \cite{Marcus89}-\cite{Ernst98}, 
FitzHugh-Nagumo (FN) \cite{Frank01b}-\cite{Campbell01},
Hindmarsh-Rose (HR) \cite{Rosenblum04},
and Hodgkin-Huxley (HH) models \cite{Foss96}
\cite{Pakdaman96}\cite{Park96}\cite{Hasegawa00b}.
Exposed behaviors due to time delays are the multistability and
bifurcation leading to chaos.
 
There are two difficulties in studying combined effects of 
noise and delay in brains.
One is that the system is usually described
by {\it nonlinear} SDDEs, 
which are generally more difficult than
linear SDDEs.
Dynamics of individual neurons includes a variety of 
voltage dependent ionic channels which can be described by 
nonlinear DEs of Hodgkin-Huxley-type models, 
or of reduced neuron models such as IF, FN and HR models.
The other difficulty is that a small cluster of cortex 
consists of thousands of similar neurons. 
For a study of dynamics of noisy neuron
ensembles with time-delayed couplings,
we have to solve high-dimensional nonlinear SDDEs, which
have been studied by
direct simulations (DSs) \cite{Kim97} \cite{Borisyuk02}
and by analytical methods like FPE \cite{Zorzano03}. 
Simulations for large-scale neuron ensembles have been made 
mostly by using IF, FN, HR and phase models.
Since the time to simulate networks by conventional methods 
grows as $N^2$ with $N$, the size of the ensemble,
it is rather difficult to simulate realistic neuron clusters.
Although FPE is a powerful method in dealing with the stochastic DE, 
a simple FPE application to SDDE fails because of 
its non-Markovian property \cite{Duil99}\cite{Frank01}.
  
In a preceding paper \cite{Hasegawa04} (which is referred to as I hereafter),
the present author has developed an augmented moment method (AMM)
for SDDE, employing a semi-analytical dynamical 
mean-field approximation (DMA) theory \cite{Hasegawa03}\cite{Hasegawa03b}.
In I, AMM is applied to an ensemble described by
the delay Langevin model, transforming the original
$N$-dimensional SDDEs to infinite-dimensional DEs which are
terminated at finite level $m$ in the level-$m$ AMM (AMM$m$).
Model calculations in I with changing the level $m$ have shown
that calculated results converge at a fairly small $m$.
Actually results obtained by AMM6 are
in good agreement with those by DSs
for linear and nonlinear Langevin ensembles.  
It has been demonstrated in I that AMM may be a useful tool
in discussing dynamics and synchronization of ensembles
described by SDDEs.

It is the purpose of the present paper to apply AMM to
FN neuron ensembles with time-delayed couplings.
In the next Sec. II, we apply our AMM theory
to nonlinear SDDEs of $N$-unit FN neuron ensembles, 
in order to get the infinite-dimensional deterministic DEs for
the correlation functions of local and global variables.
Infinite-dimensional recursive DEs are terminated at
the finite level $m$ in AMM$m$. 
In Sec. III we report model calculations, 
showing that results of our AMM are
in good agreement with those of DSs.
Section IV is devoted to 
conclusions and discussions.

\section{FN Neuron ensemble}

\subsection{Adopted model and method}

Dynamics of a neuron ensemble consisting of $N$-unit
FN neurons ($N \geq 2$), 
is described by the $2N$-dimensional 
nonlinear SDDEs given by 
\begin{eqnarray}
\frac{dx_{1i}(t)}{dt} &=& F[x_{1i}(t)]- c x_{2i}(t) 
+ \left( \frac{1}{N-1}\right) 
\;\sum_{j (\neq i)} w_{ij}\: G(x_{1j}(t-\tau_{ij}))
+ \xi_i(t)+I^{(e)}(t), \\
\frac{d x_{2i}(t)}{dt} &=& b x_{1i}(t) - d x_{2i}(t)+e,
\hspace{3cm}\mbox{($i=1-N$)}
\end{eqnarray}
where $F[x(t)]=k\: x(t)\: [x(t)-h]\: [1-x(t)]$, 
$k=0.5$, $h=0.1$, $b=0.015$, $c=1.0$, $d=0.003$ and $e=0$ 
\cite{Hasegawa03}\cite{Rod96}, and
$x_{1i}$ and $x_{2i}$ denote the fast (voltage)
and slow (recovery) variables, respectively.
The third term in Eq. (1) stands for interactions 
with the uniform couplings of $w_{ij}=w$ and
delay times of $\tau_{ij}=\tau$, 
and the sigmond function $G(x)$ given by
$G(x)=1/(1+exp[-(x-\theta)/\alpha])$,
$\theta$ and $\alpha$ denoting the threshold and
the width, respectively \cite{Note6}. 
The all-to-all couplings have been widely employed in theoretical studies.
The assumed constant delay may be justified in certain neural
networks \cite{Salami03}.
The fourth term of Eq. (1), $\xi_{i}(t)$, denotes the
Gaussian white noise given by
$<\xi_i(t)>=0$ and
$<\xi_i(t)\:\xi_j(t')> = \beta^2 \delta_{ij} \delta(t-t')$
where $\beta$ denotes the magnitudes of independent noises
and the bracket $<\cdot>$ the stochastic average \cite{Note2}.
The last term in Eq. (1), $I^{(e)}(t)$, denotes an external input whose
explicit form will be shown later [Eq. (31)].

We apply our AMM developed in I to FN neuron ensemble 
given by Eqs. (1) and (2),
defining global variables for the ensemble given by
\begin{eqnarray}
X_{\kappa}(t)&=&\frac{1}{N}\;\sum_{i} \;x_{\kappa i}(t), 
\hspace{1cm}\mbox{$\kappa=1,\:2$},
\end{eqnarray}
and their averages by
\begin{eqnarray}
\mu_{\kappa}(t)&=&<X_{\kappa}(t)>.
\end{eqnarray}
We define the correlation functions
between local variables, given by
\begin{eqnarray}
\gamma_{\kappa,\lambda}(t,t')&=&\frac{1}{N} \sum_i
< \delta x_{\kappa i}(t)\:\delta x_{\lambda i}(t')>, 
\hspace{1cm}\mbox{$\kappa, \:\lambda=1,\:2 $}
\end{eqnarray}
where $\delta x_{\kappa i}(t)= x_{\kappa i}(t)-\mu_{\kappa}(t)$.
Similarly we define the correlation function
between global variables, given by
\begin{eqnarray}
\rho_{\kappa,\lambda}(t,t')
&=&< \delta X_{\kappa}(t)\;\delta X_{\lambda}(t')  >, \\
&=&\frac{1}{N^2} \sum_i \sum_j
< \delta x_{\kappa j}(t)\:\delta x_{\lambda i}(t')>, 
\end{eqnarray}
where $\delta X_{\kappa}(t) 
= X_{\kappa}(t)-\mu_{\kappa}(t)$.
Conventional variances and covariances are given  by Eqs. (5)-(7)
with $t=t'$, for which
the symmetry relations: 
$\gamma_{1,2}(t,t)=\gamma_{2,1}(t,t)$ and
$\rho_{1,2}(t,t)=\rho_{2,1}(t,t)$, are hold.
It is noted that $\gamma_{\kappa,\nu}(t,t)$ ($\kappa, \;\lambda=1,\;2$) 
expresses the spatial average of fluctuations
in local variables of $x_{\kappa i}$ 
while $\rho_{\kappa,\nu}(t,t)$
denotes fluctuations in global variables of $X_{\kappa}$.

After our previous studies 
\cite{Hasegawa04,Hasegawa03,Hasegawa03b},
we have assumed that the noise intensity $\beta$ is weak 
and that the distribution of state variables takes the Gaussian form
concentrated near the means of ($\mu_1,\:\mu_2$). 
The second assumption is justified from numerical calculations 
for single FN \cite{Tuckwell98,Tanabe01}
and HH neurons \cite{Tanabe99,Tanabe01a}.
We will obtain infinite-order equations of motions for means,
variance and covariances defined by Eqs. (5)-(7).
They will be terminated at the level $m$ in AMM$m$.
Readers who are not interested in mathematical details, may skip 
to Sec. IIC.

\subsection{Equations of motions}

After some manipulations, we get DEs for 
$\mu_{\kappa}(t)$, $\gamma_{\kappa,\nu}(t,t)$
and $\rho_{\kappa,\nu }(t,t)$ ($\kappa, \nu=1,2$) given by
(for details see appendix A)
\begin{eqnarray}
\frac{d \mu_1(t)}{d t}&=&f_0(t) + f_2(t) \gamma_{1,1}(t,t) -c \mu_2(t) 
+ w\; u_0(t-\tau)+I^{(e)}(t), \\
\frac{d \mu_2(t)}{d t}&=& b \mu_1(t) - d \mu_2(t) +e, \\
\frac{d \gamma_{1,1}(t,t)}{d t}&=& 2 [ a(t) 
\gamma_{1,1}(t,t)- c \gamma_{1,2}(t,t) ] 
+  2w u_1(t-\tau) \:\zeta_{1,1}(t,t-\tau)
+\beta^2, \\
\frac{d \gamma_{2,2}(t,t)}{d t}
&=& 2 [ b \gamma_{1,2}(t,t)- d \gamma_{2,2}(t,t) ],  \\
\frac{d \gamma_{1,2}(t,t)}{d t}&=& b \gamma_{1,1}(t,t)
+ [a(t)-d] \gamma_{1,2}(t,t) 
- c \gamma_{2,2}(t,t)
+ w u_1(t-\tau) \:\zeta_{2,1}(t,t-\tau), \\
\frac{d \rho_{1,1}(t,t)}{d t}&=& 
2 [ a(t) \rho_{1,1}(t,t) - c \rho_{1,2}(t,t) ] 
+2  w u_1(t-\tau)  \rho_{1,1}(t,t-\tau)
+ \frac{\beta^2}{N}, \\
\frac{d \rho_{2,2}(t,t)}{d t}&=& 2 [b \rho_{1,2}(t,t)- d \rho_{2,2}(t,t)],  \\
\frac{d \rho_{1,2}(t,t)}{d t}&=& b \rho_{1,1}(t,t)
+ [a(t)-d] \rho_{1,2}(t,t) 
- c \rho_{2,2}(t,t) 
+  w u_1(t-\tau) \rho_{2,1}(t,t-\tau),  
\end{eqnarray}
with
\begin{eqnarray}
a(t)&=& f_1(t)+3 f_3(t) \gamma_{1,1}(t,t), \\
u_0(t)&=&g_0(t)+g_2(t) \gamma_{1,1}(t,t),\\
u_1(t)&=&g_1(t)+3 g_3(t) \gamma_{1,1}(t,t),\\
\zeta_{\kappa,\nu}(t,t')
&=&\left( \frac{1}{N-1} \right)
[N \rho_{\kappa,\nu}(t,t')-\gamma_{\kappa,\nu}(t,t')],
\end{eqnarray}
where $f_{\ell}(t)=(1/\ell \:!) F^{(\ell)}(\mu_1(t))$ and 
$g_{\ell}(t)=(1/\ell \:!) G^{(\ell)}(\mu_1(t))$.
Equations (8)-(15) include the higher-order terms of 
$\gamma_{\kappa,\nu}(t,t-\tau)$ and 
$\rho_{\kappa,\nu}(t,t-\tau)$, whose
equations of motions are given by ($m \ge 1$)
\begin{eqnarray}
\frac{d \gamma_{1,1}(t,t-m\tau)}{d t}&=& 
[a(t) + a(t-m\tau)] 
\gamma_{1,1}(t,t-m\tau) 
- c [\gamma_{1,2}(t,t-m\tau) + \gamma_{2,1}(t,t-m\tau)]\nonumber \\
&+& w[ u_1(t-\tau)\:\zeta_{1,1}(t-\tau,t-m\tau) \nonumber \\
&+&  u_1(t-(m+1)\tau)\:\zeta_{1,1}(t,t-(m+1)\tau) ]+\beta^2\:\Delta(m \tau), \\
\frac{d \gamma_{2,2}(t,t-m\tau)}{d t}
&=& b [\gamma_{1,2}(t,t-m\tau)+ \gamma_{2,1}(t,t-m\tau)]
- 2d \gamma_{2,2}(t,t-m\tau),  \\
\frac{d \gamma_{1,2}(t,t-m\tau)}{d t}&=& b \gamma_{1,1}(t,t-m\tau)
+ [a(t)-d] \gamma_{1,2}(t,t-m\tau) 
- c \gamma_{2,2}(t,t-m\tau) \nonumber \\
&+&  w u_1(t-\tau)\:\zeta_{1,2}(t-\tau,t-m\tau),  \\
\frac{d \gamma_{2,1}(t,t-m\tau)}{d t}&=& b \gamma_{1,1}(t,t-m\tau)
+ [a(t-m\tau)-d] \gamma_{2,1}(t,t-m\tau) 
- c \gamma_{2,2}(t,t-m\tau) \nonumber \\
&+&  w u_1(t-(m+1)\tau)\:\zeta_{2,1}(t,t-(m+1)\tau),  \\
\frac{d \rho_{1,1}(t,t-m\tau)}{d t}&=& 
[a(t) + a(t-m\tau)] 
\rho_{1,1}(t,t-m\tau) 
- c [\rho_{1,2}(t,t-m\tau)+\rho_{2,1}(t,t-m\tau)]  \nonumber \\
&+& w [u_1(t-\tau)\rho_{1,1}(t-\tau,t-m\tau) \nonumber \\
&+& u_1(t-(m+1)\tau) \rho_{1,1}(t,t-(m+1)\tau) ]
+\left( \frac{\beta^2}{N} \right) \:\Delta(m\tau), \\
\frac{d \rho_{2,2}(t,t-m\tau)}{d t}
&=& b [\rho_{1,2}(t,t-m\tau)+ \rho_{2,1}(t,t-m\tau)]
- 2 d \rho_{2,2}(t,t-m\tau),  \\
\frac{d \rho_{1,2}(t,t-m\tau)}{d t}&=& b \rho_{1,1}(t,t-m\tau)
+ [a(t)-d] \rho_{1,2}(t,t-m\tau) 
- c \rho_{2,2}(t,t-m\tau) \nonumber \\
&+&  w  u_1(t-\tau) \rho_{1,2}(t-\tau,t-m\tau),  \\
\frac{d \rho_{2,1}(t,t-m\tau)}{d t}&=& b \rho_{1,1}(t,t-m\tau)
+ [a(t-m\tau)-d] \rho_{2,1}(t,t-m\tau) 
- c \rho_{2,2}(t,t-m\tau) \nonumber \\
&+&  w  u_1(t-(m+1)\tau) \rho_{2,1}(t,t-(m+1)\tau),
\end{eqnarray}
where $\Delta(x)=1$ for $x=0$ and 0 otherwise.

\subsection{Summary of our method}

The original two-dimensional SDDE given by Eqs. (1) and (2)
are transformed to infinite-dimensional
deterministic DDEs given by Eqs. (8)-(15) and (20)-(27),
which are due to non-Markovian property of SDDE. 
It is, however, impossible to simultaneously 
solve these infinite-order recursive equations.
We will adopt the level-$m$ AMM (AMM$m$) in which
the recursive DEs are terminated at the finite level $m$, as
\begin{eqnarray}
\gamma_{\kappa,\nu}(t,t-(m+1)\tau) 
&=&\gamma_{\kappa,\nu}(t,t-m\tau), \\ 
\rho_{\kappa,\nu}(t,t-(m+1)\tau) 
&=& \rho_{\kappa,\nu}(t,t-(m+1)\tau), \\ 
g_1(t-(m+1)\tau) &=& g_1(t-m\tau),
\end{eqnarray}
leading to $8(m+1)$-dimensional DEs.
In the following Sec. III, we will examine 
AMM$m$, performing calculations
with changing $m$, in order to show that AMM5
may yield results in fairy good agreement 
with those of DS [Fig. 5(b)].
In the limit of $\tau=0$,
Eqs. (20)-(27) reduce to Eqs. (10)-(15),
then Eqs. (8)-(15) agree with Eqs. (20)-(27) 
in Ref. \cite{Hasegawa03} for FN neurons ensembles
without delays \cite{Note6}.

Model calculations will be reported in
the following Sec. III.
DSs have been performed for
$2 N$ DEs given by  Eqs. (1) and (2) 
by using the fourth-order Runge-Kutta method with
a time step of 0.01.
Initial values of variables at $t \in (-\tau, 0]$ are 
$x_i(t)=y_i(t)=0$ for $i=1$ to $N$. 
DS results
are the average of 100 trials otherwise noticed.
AMM calculations have been performed for
Eqs. (8)-(30) by using also
the fourth-order Runge-Kutta method with
a time step of 0.01.
Initial values are
$\mu_1(t)=\mu_2(t)=0$ at $t \in [-\tau, 0]$, and 
$\gamma_{\kappa,\nu}(t,t')=\rho_{\kappa,\nu}(t,t')=0$
$t \in [-\tau, 0]$ or $t' \in [-\tau, 0]$ ($t \geq t'$).
All calculated quantities are dimensionless. 

\section{Model calculations}

\subsection{Effects of coupling ($w$) and delay ($\tau$)}

In this study, we pay our attention to
the response of the FN neuron ensembles
to a single spike input of $I^{(e)}(t)$
given by \cite{Hasegawa03}
\begin{equation}
I^{(e)}(t)
=A \;\Theta(t-t_{in})\; \Theta(t_{in}+T_w-t),
\end{equation}
where $\Theta(x)=1$ for $x>0$ and 0 otherwise,  
$A$ stands for the magnitude, $t_{in}$ the input time and 
$T_w$ the spike width.
We have adopted the same parameters of 
$A=0.10$, $t_{in}=100$ and $T_w=10$ as in Ref. \cite{Hasegawa03}. 
Parameter values of $w$, $\tau$, $\beta$
and $N$ will be explained shortly.

When an input spike given by Eq. (31) is applied,
the oscillation may be triggered 
when model parameters are appropriate.  
The $w$-$\tau$ phase diagram showing the oscillating (OSC)
and non-oscillating (NOSC) states is depicted in Fig. 1, which is
calculated for $\beta=0$ and $N=10$.
In the case of $\beta=0.01$, for example, 
the OSC region is slightly shrunk compared to that for $\beta=0$,
as will be shortly discussed [Figs. 5(a) and 5(b)].
The $w$-$\tau$ phase is
separated by two boundaries in positive- and negative-$w$ regions.
Circles in Fig. 1 express pairs
of $w$ and $\tau$ adopted for calculations
to be shown in Figs. 2 and 3.
Along the horizontal, dashed line in Fig. 1,
the $w$ value is continuously changed in calculations to be shown 
in Figs. 4(a) and 4(b). 

In order to monitor the emergence of the oscillation, 
we calculate the quantity: 
\begin{equation}
\sigma_o=\overline{O(t)}
=\frac{1}{t_2-t_1} \int_{t_1}^{t_2}\: dt \;O(t),
\end{equation}
with
\begin{eqnarray}
O(t)&=& \frac{1}{N} 
\sum_i [<x_i(t)^2>-<x_i(t)>^2], \\
&=& \mu(t)^2-\mu(t)^2+\gamma_{1,1}(t),
\end{eqnarray}
which becomes finite in the oscillation state but
vanishes in the non-oscillating state,
the overline denoting the temporal average
between $t_1$ (=2000) and $t_2$ (=4000).

The synchrony within ensembles is measured by
\cite{Hasegawa04}\cite{Hasegawa03}
\begin{equation}
\sigma_s = \overline{S(t)},
\end{equation}
with
\begin{eqnarray}
S(t)&=& \left( \frac{N\rho_{1,1}(t,t)/\gamma_{1,1}(t,t)-1}{N-1} \right),
\end{eqnarray}
which is 0 and 1 for completely asynchronous and synchronous states,
respetively. 

We have calculated
time courses of $\mu_1(t)$, $\gamma_{1,1}(t,t)$, $\rho_{1,1}(t,t)$
and $S(t)$,
whose results are depicted in Figs. 2(a)-2(l),
solid and dashed curves denoting results of AMM and
DS, respectively.

For $\tau=0$, an output spike of $\mu_1(t)$ fires
after an applied input which is plotted 
at the bottom of Fig. 2(a) [and also of 2(e) and 2(i)].
It is noted that state variables are randomized when an input
spike is applied at $t=100$ because independent noises have been added 
since $t=0$.
Figures 2(b) and 2(c) show $\gamma_{1,1}$ and $\rho_{1,1}$
for $\tau=0$, respectively. 
The synchronization ratio $S(t)$ for $\tau=0$ shown
in Fig. 2(d) has an appreciable magnitude:
its maximum values calculated in AMM are 0.038 and 0.077 at $t=107$
and 123, respectively. 
Figure 2(e) shows that
when a delay of $\tau=20$ is introduced, 
an input signal leads to a spike output with
an additional, small peak in $\mu_1$ at $t= 133$.
Figures 2(f) and 2(g) show that
although a peak of $\gamma_{1,1}$ for $\tau=20$ becomes
larger than that for $\tau=0$,
a peak of $\rho_{1,1}$ is decreased by an introduced delay. 
Maximum values of $S(t)$ calculated by AMM
are 0.154 and 0.130 at $t=126$ and 140, respectively, for $\tau=20$.
We note from Fig. 2(i) that
for a larger $\tau=60$, an input spike triggers 
an autonomous oscillation with a period of about 65. 
Peaks in $\gamma_{1,1}$, $\rho_{1,1}$ and $S$ are progressively
increased with increasing $t$ as shown in Figs. 2(j), 2(k) and 2(l):
peaks of $\gamma_{1,1}$, $\rho_{1,1}$ and $S$
saturate at $t \simg 1200$ with the values of 0.00253, 0.00014
and 0.098, respectively, in AMM calculations.
We note in Figs. 2(a)-2(l) that results of $\mu_1$ obtained by AMM and DS
are indistinguishable, and 
that AMM results of $\gamma_{1,1}$, $\rho_{1,1}$ and $S$ are
in fairly good agreement with those of DSs. 

Figure 1 shows that
although the obtained NOSC-OSC phase is nearly symmetric 
with respect to the $w=0$ axis, it is not in the strict sense.
Actually the property of the oscillation for inhibitory couplings ($w < 0$) is
different from that for excitatory couplings ($w > 0$).
Figures 3(a) and 3(b) show autonomous oscillations for $w=0.1$ and $w=-0.1$,
respectively, with $\tau=60$, $\beta=0.01$ and $N=10$.
The period of the oscillation $T$ is given by
$T=\tau+\tau_i$ where $\tau_i$ denotes the intrinsic delay
for firings.
For inhibitory feedback with negative $w$,
FN neurons fire with the rebound process,
which requires a larger $\tau_i$ for firing than for excitatory 
feedback with positive $w$.
Then the period of $T = 86$ for autonomous oscillation 
with the negative $w$ 
becomes larger than that of $T = 65$ with the positive $w$.   

By changing the $w$ value along the horizontal, dashed line in Fig. 1, 
we have calculated the $w$ dependence of $\sigma_o$ 
and $\sigma_s$, whose results 
are plotted in Figs. 4(a) and 4(b), respectively,
for $\beta=0.0001$ and 0.01.
The oscillation emerges for $w \simg 0.058$ or $w \siml -0.063$ 
with $\beta=0.0001$, while with $\beta=0.01$ it occurs
for $w \simg 0.060$ or $w \siml -0.070$.
The transition from NOSC to OSC states
is of the first order because
$\sigma_o$ is abruptly increased 
at the critical coupling of $w=w_c$, where $\sigma_s$ has a narrow peak.
In contrast, the relevant NOSC-OSC transition 
in the nonlinear Langevin model is of the second order \cite{Hasegawa04}.

We have investigated, in more detail,
the $w$ dependence of $\sigma_o$ and $\sigma_s$ near the transition region
of $0.05 \leq w \leq 0.07$,
which is sandwiched by vertical, dashed lines in Figs. 4(a) and 4(b),
results for $\beta=0.0001$ and $\beta=0.01$ being plotted 
in Figs. 5(a) and 5(b), respectively. 
Figure 5(a) shows
that the critical $w$ value for the NOSC-OSC transition 
is $w_c \simeq 0.0579$ for $\beta=0.0001$ both in DS and AMM5.
When we adopt AMM1, we get the result showing the
NOSC-OSC transition at $w \sim 0.6$, although we cannot get
solutions for $0.0586 < w < 0.060$.
With the use of AMM2, 
we get the transition at $w \sim 0.058$,
though solutions are not obtainable
for $0.0580 < w < 0.0582$.
We have noted that AMM$m$ converges at the level $m = 3$, 
above which calculated results are almost identical.
Figure 5(b) shows that the critical value of $w_c$ 
for $\beta=0.01$ is 0.0600 in DS and 0.0607 in AMM5.
For $m=1$, 2 and 3, the NOSC-OSC transition occurs 
at $w =$ 0.0644, 0.0609 and 0.0807,
respectively: $w_c$ for $m=3$ approaches that for $m=5$
(in what follows results of AMM5 will be reported).
It is interesting to note in Figs. 5(a) and 5(b) that 
the synchrony $\sigma_s$ shows 
{\it fluctuation-induced} enhancement
at the NOSC-OSC transition. This is due to an increase in the ratio of
$\rho_{1,1}(t,t)/\gamma_{1,1}(t,t)$ in Eq. (36) although 
both $\rho_{1,1}(t,t)$ and $\gamma_{1,1}(t,t)$ are increased
at the NOSC-OSC transition.
Similar phenomenon has been reported 
in the nonlinear Langevin model \cite{Hasegawa04} and
in heterogeneous systems
in which the oscillation emerges when the degree of the heterogeneity
exceeds the critical value \cite{Cartwright00}\cite{Boschi01}. 

\subsection{Effects of noise ($\beta$)}

Comparing Fig. 5(b) with Fig. 5(a),
we note that when the noise intensity 
is increased form $\beta=0.0001$ to $\beta=0.01$,
the critical $w_c$ value for the NOSC-OSC transition is increased:
$w_c$=0.0579 (0.0579) for $\beta=0.001$ and
$w_c$=0.0600 (0.0607) for $\beta=0.01$ in DS (AMM). 
Figure 6(a) shows the $\beta$ dependence of
$\sigma_o$ and $\sigma_s$ for $\tau=60$, $w=0.06$ and $N=10$.
$\sigma_o$ is rapidly decreased at $\beta \sim \beta_c$ where
$\sigma_s$ has a broad peak: $\beta_c$ is about 0.01 in DS
while it is about 0.0075 in AMM.
Figure 6(b) shows that the similar $\beta$ dependence of 
$\sigma_o$ and $\sigma_s$
is obtained also for a larger $w=0.062$, for which
$\beta_c \sim 0.015$ in DS
and $\beta_c \sim 0.014$ in AMM.
A suppression of the oscillation by noises is realized
in the Langevin model \cite{Hasegawa04} and in
some calculations for systems with heterogeneity \cite{Boschi01}, 
although the noise-induced oscillation is 
reported in Refs. \cite{Zorzano03}\cite{Hu00}\cite{Vries01}. 
In particular, Zorzano and V\'{a}zquez \cite{Zorzano03} (ZV) 
showed the noise-induced
oscillation in FN neuron ensembles ($N=\infty$) with time delays
by using FPE method.  
The difference between ZV's results and ours
may be due to the difference in the adopted FN model
and/or ensemble size.
In order to get some insight on this issue, 
we have performed AMM calculations for our FN model
with larger ensemble sizes of $N=100$ and 1000, 
and obtained again a suppression of the oscillation 
by noises \cite{Note3}. 
It is not clear for us how ZV took into account 
the non-Markovian property of SDDE 
within their FPE method \cite{Duil99}\cite{Frank01}.

\subsection{Effects of size ($N$)}

The $N$ dependence of $\sigma_o$ and $\sigma_s$
for $\beta=0.01$, $w=0.06$ and $\tau=60$ is shown in Fig. 7 where 
open circles (squares) express 
$\sigma_o$ ($\sigma_s$) in DS, and
where thin (bold) solid curves denote $\sigma_o$ ($\sigma_s$) in AMM.
It is shown that with increasing the size of ensemble,
$\sigma_o$ is gradually increased at $N \sim N_c$ where
$\sigma_s$ has a broad peak,
the critical dimension being $N_c \sim $10 in DS
and $N_c \sim $ 100 in AMM. 
Results of our AMM calculations
are qualitatively similar to those of DS
although calculated $N_c$ values are different
between the two methods.

\section{Conclusions and Discussions}

In Sec. II, we have obtained the infinite-dimensional 
ordinary differential equations.
It is, however, possible to get expressions
given by partial differential equations (PDEs) if we define
the correlation functions:
\begin{eqnarray}
C_{\kappa,\lambda}(t,z)&=& 
\frac{1}{N} \sum_i <\delta x_{\kappa i}(t) 
\:\delta x_{\lambda i}(t-z)>,\\
D_{\kappa,\lambda}(t,z)&=& 
<\delta X_{\kappa}(t) \:\delta X_{\lambda}(t-z)>,
\end{eqnarray}
introducing a new variable $z$ [see Eqs. (5) and (6)].
For example, PDEs for $C_{1,1}(t,z)$ are given by
\begin{eqnarray}
\frac{\partial C_{1,1}(t,0)}{\partial t}
&=& 2[a C_{1,1}(t,0) - c C_{1,2}(t,0)] 
+ 2 w u_1(t-\tau) E_{1,1}(t,\tau) + \beta^2, \\
\left( \frac{\partial}{\partial t}
+\frac{\partial}{\partial z} \right) C_{1,1}(t,z)
&=&a C_{1,1}(t,z)- c C_{1,2}(t,z)\nonumber \\
&+& w u_1(t-\tau) E_{1,1}(t-\tau,z-\tau), 
\hspace{2cm}\mbox{for $z > 0$}
\end{eqnarray}
where $E_{1,1}(t,z)=[N D_{1,1}(t,z)-C_{1,1}(t,z)]/[N-1]$.
It is noted that Eqs. (39) and (40) correspond 
to Eqs. (10) and (20), respectively.
Then we have to solve PDEs including $\mu_{\kappa}(t)$,
$C_{\kappa,\lambda}(t,z)$
and $D_{\kappa,\lambda}(t,z)$
with a proper boundary condition in the $(t, z)$ space.
A similar PDE approach has been adopted in Ref. \cite{Frank03}
for an analysis of the stationary solution of the
linear Langevin equation with delays.
In an earlier stage of this study, we pursued the 
PDE approach. We realized, however, 
from the point of computer programming that
the use of the ordinary DEs given in AMM
is more tractable than that of PDEs.

Our calculations have shown that
FN neuron ensembles with delays exhibit the multistability
when model parameters such as
$w$, $\tau$, $\beta$ and $N$ are varied.
The multistability is the common property of the
system with time delay.
Actually the nonlinear Langevin ensembles discussed
in I also show the multistability:
the $w-\tau$ phase diagram of FN ensembles
shown in Fig. 1 is similar 
to that of the Langevin ensembles shown in Fig. 6 of I.
In either case,
{\it fluctuation-induced} synchronization is realized 
near the transition between OSC and NOSC states.
These results imply that 
the oscillating, highly synchronous
state may be realized in ensembles for smaller couplings
with a proper delay 
than with no delays.
This is consistent with the recent result of 
Ref. \cite{Dhamala04}, where the importance
of delays is stressed for the long-range synchronization with
low coupling strength.

In summary, we have discussed dynamics of FN neuron ensembles
with delays by using a semi-analytical method
developed in I.
Our method has a limitation of weak noises but it is
free from the magnitude of delay times. 
This is complementary to the
small-delay approximation \cite{Duil99}, whose application
to FN neuron ensembles with delays is discussed in appendix C.
For FN ensembles to show the oscillation, we have to adopt
an appreciable magnitude of delay ($\tau \simg 20$), for which 
SDA method cannot be employed. 
In this study we have discussed only the case of 
a single spike input.
Our method may be, however, applicable to arbitrary inputs
such as periodic spike trains and Poisson spikes,
as was made for HH neuron ensembles (without delays) 
\cite{Hasegawa03b}.  
Although results calculated by our method are in fairly good agreement
with those obtained by DC, 
the quantitative analytical theory is still lacking.
In this study, we have assumed regular
couplings ($w_{ij}=w$)
and uniform time delays ($\tau_{ij}=\tau$). In real systems, however,
couplings are neither regular nor random, and
time delays are nonuniform with a variety of dendrite radius
and length.
It is interesting to include these properties by extending
our approach, which is in progress and will be reported
in a future paper.

\section*{Acknowledgments}
This work is partly supported by
a Grant-in-Aid for Scientific Research from the Japanese 
Ministry of Education, Culture, Sports, Science and Technology.  


\appendix

\section{Derivation of Eqs. (8)-(15)}

We express Eqs. (1) and (2) in a Taylor expansion
of $\delta x_i \:(=\delta x_{1i})$ 
and $\delta y_i \:(=\delta x_{2i})$ 
up to the third-order terms to get
\begin{eqnarray}
\frac{d \delta x_i(t)}{d t}
&=& f_1(t) \delta x_i(t)+f_2 [\delta x_i(t)^2-\gamma_{1,1}(t,t)]
+ f_3(t) \delta x_i(t)^3 - c \delta y_i(t) \nonumber \\
&+& \xi_i(t) + \delta I_i^{(c)}(t-\tau), \\
\frac{d \delta y_i(t)}{d t}&=& b \delta x_i(t) - d \delta y_i(t),
\end{eqnarray}
with
\begin{eqnarray}
\delta I_i^{(c)}(t)
&=&  w \left( \frac{g_1(t)}{N-1} \sum_{j(\neq i)} \delta x_j(t)
+  \frac{g_2(t)}{N-1} \sum_{j(\neq i)} [\delta x_j(t)^2 -\gamma_{1,1}]
+  \frac{g_3(t)}{N-1} \sum_{j(\neq i)} \delta x_j(t)^3 \right), 
\end{eqnarray}
where 
$f_{\ell}(t)=(1/\ell \:!) F^{(\ell)}(\mu_1(t))$ and 
$g_{\ell}(t)=(1/\ell \:!) G^{(\ell)}(\mu_1(t))$.
Averages of Eqs. (A1) and (A2) with Eqs. (3) and (4) yield
DEs for means of $d \mu_1/d t$ and $d \mu_2/d t$ [Eq. (8)].
DEs for variances and covariances may be obtained 
by using the equations of motions of 
$\delta x_i$ and $\delta y_i$.
For example, DE for $d \gamma_{1,2}(t,t)/d t$ is given by
\begin{equation}
\frac{d \gamma_{1,2}(t,t)}{d t} = \frac{1}{N} \sum_i
< \left(\frac{d \delta x_i(t)}{d t}\right) \;\delta y_i(t)
+\delta x_i(t) \;\left(\frac{d \delta y_i(t)}{d t}\right) >,
\end{equation} 
which leads to Eq. (12). 
DEs for other variances and covariances are similarly obtained.

\section{Derivation of Eqs. (20) and (27)}

In the process of calculations of Eqs. (8)-(15), we get new
correlation functions given by
\begin{eqnarray}
S_{1}(t_1,t_2)&=&\frac{1}{N} \sum_i < \delta x_i(t_1)\;\xi_i(t_2) >, \\
S_{2}(t_1,t_2)&=&\frac{1}{N} \sum_i <\delta y_i(t_1) \;\xi_i(t_2)>, 
\end{eqnarray}
where $\delta x_i=\delta x_{1i}$, $\delta y_i=\delta x_{2i}$,
$t_1=t$ and $t_2=t-m\tau$, or $t_1=t-m\tau$ and $t_2=t$.
We will evaluate them by using 
DEs for $\delta x_i(t)$ and $\delta y_i(t)$,
which are linearized from Eqs. (A1)-(A3):
\begin{eqnarray}
\frac{d \delta x_i(t)}{d t}&=& a(t) \delta x_i(t) - c \delta y_i(t) 
+\left( \frac{w}{N-1} \right)
\sum_{j (\neq i)} g_1(t-\tau) \delta x_j(t-\tau)
+ \xi_i(t), \\
\frac{d \delta y_i(t)}{d t}&=& b \delta x_i(t) - d \delta y_i(t),
\end{eqnarray}
where $a(t)=f_1(t)+3 f_3(t) \gamma_{1,1}(t,t)$.
Neglecting the $t$ dependence in $a(t)$,
we get formal solutions of Eqs. (B3) and (B4)
given by
\begin{eqnarray}
\delta x_i(t)&=& \left( \frac{A+d}{A-B} \right) 
\int^t ds 
\;{\rm exp}^{(t-s) A} 
[\left( \frac{w}{N-1} \right) 
\sum_{j (\neq i)} g_1(s-\tau) \delta x_j(s-\tau)
+ \xi_i(s)]
\nonumber \\
&-& \left( \frac{B+d}{A-B} \right) 
\int^t ds 
\;{\rm exp}^{(t-s) B} 
[\left( \frac{w}{N-1} \right)
\sum_{j (\neq i)} g_1(s-\tau) \delta x_j(s-\tau)
+ \xi_i(s)], \\
\delta y_i(t)&=& \left( \frac{b}{A-B} \right) 
\int^t ds 
\;{\rm exp}^{(t-s)A}
[\left( \frac{w}{N-1} \right) 
\sum_{j (\neq i)} g_1(s-\tau) \delta x_j(s-\tau)
+ \xi_i(s)]
\nonumber \\
&-&\left( \frac{b}{A-B} \right) \int^t ds 
\;{\rm exp}^{(t-s)B} \;
[\left( \frac{w}{N-1} \right) 
\sum_{j (\neq i)} g_1(s-\tau) \delta x_j(s-\tau)
+ \xi_i(s)],
\end{eqnarray}
where $A$ and $B$ are roots of the equation given by
$z^2-(a-d)\:z-a\:d+b\:c=0$.
By using the method of steps in Ref. \cite{Frank03}, 
we obtain the step by step functions, from which we get
\begin{eqnarray}
S_1(t,t-m\tau)= S_1(t-m\tau,t)&=& 
\left( \frac{\beta^2}{2} \right) \Delta(m\tau), \\
S_2(t,t-m\tau)= S_2(t-m\tau,t)&=&0,
\end{eqnarray}
where $\Delta(x)=1$ for $x=0$ and 0 otherwise.
By using Eqs. (B7) and (B8), we get Eqs. (20)-(27).
The assumption of a neglect of the $t$ dependence in $a(t)$ 
may be justified, to some extent, 
from results calculated by
our method which are in fairly good
agreement with those by DS as reported in Sec. III.

\section{The Small-delay approximation}

When the delay $\tau$ is very small, we may adopt
the small-delay approximation (SDA)
proposed in Ref.\cite{Duil99}.
With this approximation, we first transform the 
SDDEs to stochastic non-delayed DEs,
and then to deterministic DEs with the use of DMA \cite{Hasegawa03}.
For a small $\tau$, we may expand $x_{1i}(t-\tau)$ in Eq. (1) as 
\begin{equation}
x_{1i}(t-\tau) \simeq x_{1i}(t) - \tau \frac{d x_{1i}(t)}{d t},
\end{equation}
with which Eq. (1) becomes stochastic non-delayed DEs given by
\begin{eqnarray}
&&\frac{dx_{1i}(t)}{d t} + \left( \frac{w\tau}{N-1} \right) 
\sum_{j(\neq i)} G'(x_{1j}(t)) \frac{dx_{1j}(t)}{d t} \nonumber \\
&&= F(x_{1i}) - c x_{i2} + \left( \frac{w}{N-1} \right) 
\sum_{j(\neq i)} G(x_{1j}(t)) 
+ \xi_i(t)+I^{(e)}.
\end{eqnarray}
When we apply DMA to $2N$-dimensional 
stochastic DEs given by Eqs. (2) and (C2), 
we get equations of motions for means, variances and
covariances, given by 
\begin{eqnarray}
\frac{d \mu_1(t)}{d t}&=&[1-w\tau u_1]
[ f_0(t) + f_2(t) \gamma_{1,1}(t,t) -c \mu_2(t) 
+w g_0(t)
+I^{(e)}(t) ], \\
\frac{d \mu_2(t)}{d t}&=& b \mu_1(t) - d \mu_2(t) +e, \\
\frac{d \gamma_{1,1}(t,t)}{d t}
&=& 2[a(t) \gamma_{1,1}(t,t)- c \gamma_{1,2}(t,t)
+ w u_1(t) \zeta_{1,1}(t,t)] +\beta^2
\nonumber \\
&-& 2 w \tau u_1(t) \left[ a(t)\zeta_{1,1}(t,t)- c \zeta_{1,2}(t,t)
+  \left(\frac{w u_1(t)}{N-1}\right) 
(N \rho_{1,1}(t,t)-\zeta_{1,1}(t,t)) \right], \\
\frac{d \gamma_{2,2}(t,t)}{d t}
&=& 2 [ b \gamma_{1,2}(t,t)- d \gamma_{2,2}(t,t) ],  \\
\frac{d \gamma_{1,2}(t,t)}{d t}
&=& b \gamma_{1,1}(t,t)+[a(t)-d] \gamma_{1,2}(t,t)
-c \gamma_{2,2}(t,t)+w u_1(t) \zeta_{1,2}(t,t) \nonumber \\
&&-w \tau u_1(t)\left[a(t) \zeta_{1,2}(t,t)-c \zeta_{2,2}(t,t)
+\left(\frac{w u_1(t)}{N-1}\right)
(N \rho_{1,2}(t,t)-\zeta_{1,2}(t,t)) \right], \\
\frac{d \rho_{1,1}(t,t)}{d t}&=& 
2 [1-w\tau u_1(t)] 
\left[ a(t) \rho_{1,1}(t,t) - c \rho_{1,2}(t,t) 
+ w u_1(t) \rho_{1,1}(t,t)
+ \frac{\beta^2}{2N} \right], \\
\frac{d \rho_{2,2}(t,t)}{d t}&=& 2 [b \rho_{1,2}(t,t)- d \rho_{2,2}(t,t)],  \\
\frac{d \rho_{1,2}(t,t)}{d t}
&=& b \rho_{1,1}(t,t)+[a(t)-d] \rho_{1,2}(t,t)-c \rho_{2,2}(t,t)
+w u_1(t)  \rho_{1,2}(t,t)
\nonumber \\&&
-w\tau u_1(t) [a(t) \rho_{1,2}(t,t) -c \rho_{2,2}(t,t)
+w u_1(t)  \rho_{1,2}(t,t)],  
\end{eqnarray}
where $a(t)$ and $\zeta_{\kappa,\lambda}(t,t)$
are given by Eqs. (16) and (19), respectively. 

A numerical comparison between AMM and
SDA is made in Fig. 8, where solid and chain curves
denote results of AMM and SDA, respectively.
For $\tau=0$ both methods lead to the identical result.
For small delays of $\tau=1$ and 2, results of SDA
are in fairly good agreement with those of AMM.
As the delay is increased to $\tau > 5$, however,
the discrepancy between the two methods becomes significant. 


\begin{figure}
\caption{
The $w$-$\tau$ phase diagram showing the oscillating (OSC) 
and non-oscillating (NOSC)
states for $\beta=0$ and $N=10$.
For sets of parameters of $w$ and $\tau$ marked by circles, 
time courses of $\mu(t)$, $\gamma(t,t)$, 
$\rho(t,t)$ and $S(t)$ are calculated, whose results
are shown in Figs. 2 and 3.
Along the horizontal dashed line ($\tau=60$), 
the $w$ dependence of $\sigma_o$ and $\sigma_s$ is  
calculated in Figs. 4 and 5.
}
\label{fig1}
\end{figure}

\begin{figure}
\caption{
(color online).
Time courses of $\mu_1(t)$, $\gamma_{1,1}(t)$, $\rho_{1,1}(t)$
and $S(t)$ calculated by AMM theory (solid curves) 
and DS (dashed curves)
with $A=0.10$, $\beta=0.01$, $w=0.1$ and $N=10$:
(a) $\mu_1$, (b) $\gamma_{1,1}$, (c) $\rho_{1,1}$
and (d) $S$ for $\tau=0$,
(e) $\mu_1$, (f) $\gamma_{1,1}$, (g) $\rho_{1,1}$
and (h) $S$ for $\tau=20$, and
(i) $\mu_1$, (j) $\gamma_{1,1}$, (k) $\rho_{1,1}$
and ($\ell$) $S$ for $\tau=60$.
Chain curves at bottoms of (a), (e) and (i) express
input spikes.
}
\label{fig2}
\end{figure}

\begin{figure}
\caption{
Time courses of $\mu_1(t)$ showing the oscillation
for (a) $w=0.1$ and (b) $w=-0.1$ with $\tau=60$, $\beta=0.01$ and $N=10$
calculated by AMM,
the result of (a) being shifted upwards by 2.
}
\label{fig3}
\end{figure}

\begin{figure}
\caption{
The $w$ dependence of (a) $\sigma_o$ and (b) $\sigma_s$ 
for $\beta=0.0001$ (solid curves) and $\beta=0.01$ (dashed curves)
with $\tau=60$ and $N=10$.
The region sandwiched by dashed, vertical lines
is enlarged in Figs. 5(a) and 5(b) for $\beta=0.0001$
and 0.01, respectively.
}
\label{fig4}
\end{figure}

\begin{figure}
\caption{
The $w$ dependence of $\sigma_o$ and $\sigma_s$ 
for (a) $\beta=0.0001$
and (b) $\beta=0.01$ with $\tau=60$ and $N=10$.
Thin and bold solid curves denote results of
$10\:\sigma_o$ and $\sigma_s$, respectively, in AMM, whereas
squares and circles express those 
of $10\:\sigma_o$ and $\sigma_s$,
respectively, in DS.
AMM results
with different level $m$ (=1, 2, 3 and 5) are shown.
Dotted lines are only for a guide of the eye (see text).
}
\label{fig5}
\end{figure}

\begin{figure}
\caption{
The $\beta$ dependence of $\sigma_o$ and $\sigma_s$ for 
(a) $w=0.60$ and (b) $w=0.62$
with $\tau=60$ and $N=10$. 
Thin and bold solid curves denote results of
$10\:\sigma_o$ and $\sigma_s$, respectively, in AMM whereas
squares and circles express those 
of $10\:\sigma_o$ and $\sigma_s$,
respectively, in DS. 
Dotted lines are only for a guide of the eye.
}
\label{fig6}
\end{figure}

\begin{figure}
\caption{
The $N$ dependence of $\sigma_o$ and $\sigma_s$ for 
$\beta=0.01$, $\tau=60$ and $w=0.06$. 
Thin and bold solid curves denote results of
$10\:\sigma_o$ and $\sigma_s$, respectively, in AMM, whereas
squares and circles express those 
of $10\:\sigma_o$ and $\sigma_s$,
respectively, in DS.
Dotted lines are only for a guide of the eye.
}
\label{fig7}
\end{figure}

\begin{figure}
\caption{
The time course of $\mu_1(t)$ 
calculated in AMM (solid curves) and in 
a small-delay approximation (SDA) (chain curves)
with $\beta=0.01$, $w=0.1$ and $N=10$ (see appendix C).
}
\label{fig8}
\end{figure}

\end{document}